\documentclass[a4paper,11pt]{article}
\usepackage{amssymb}
\usepackage[left=1in,right=1in,top=1in,bottom=1in]{geometry}   

\title{Observable Equivalence between General Relativity and Shape Dynamics}
\author{Tim A. Koslowski\\ \texttt{tkoslowski@perimeterinstitute.ca}\\ Perimeter Institute for Theoretical Physics\\ 31 Caroline St. N, Waterloo, Ontario, N2L 2Y5,  Canada}

\newtheorem{defi}{Definition}
\newtheorem{prop}{Proposition}

\begin{document}

\maketitle

\begin{abstract}
  In this conceptual paper we construct a local version of Shape Dynamics that is equivalent to General Relativity in the sense that the algebras of Dirac observables weakly coincide. This allows us to identify Shape Dynamics observables with General Relativity observables, whose observables can now be interpreted as particular representative functions of observables of a conformal theory at fixed York time. An application of the observable equivalence of General Relativity and Shape Dynamics is to define the quantization of General Relativity through quantizing Shape Dynamics and using observable equivalence. We investigate this proposal explicitly for gravity in 2+1 dimensions.
\end{abstract}

\section{Introduction}

Shape Dynamics \cite{Gomes:2010fh,Gomes:2011zi,Koslowski:2011jg} is a theory of gravity based on spatial conformal symmetry rather than spacetime covariance that turns out to be dynamically equivalent to General Relativity, which means that the physical trajectories of Shape Dynamics and General Relativity coincide. This implies that Shape Dynamics has the appealing feature that all local constraints are linear in momenta and should thus be quantizable as vector fields on the ADM configuration, i.e. the space of Riemannian metrics. This improvement (compared to the ADM formulation of gravity) is however accompanied by a nonlocal Hamiltonian which seems to obstruct quantization. This nonlocality turns out to be imposed by conformal symmetry and dynamical equivalence with General Relativity. It is the purpose of this paper to propose a remedy to this problem by splitting the equivalence of General Relativity and Shape Dynamics into two parts that are naturally separated in a Wheeler-DeWitt quantization approach, where ultimately only Dirac observables (or perennials in Kucha\v{r}'s notation\cite{Kuchar:perennial}) are considered. This split, combined with the framework of equivalence of gauge theories, suggests a particular approach to the problem of time in canonical quantum gravity\footnote{We do however not attempt to propose a solution to the problem of time in this paper.}.

Canonical quantization can be split into (a) the construction of an observable algebra $\mathfrak A$, (b) the construction of a Hilbert space representation $(\mathcal H,\pi)$ of $\mathfrak A$ on $\mathcal H$ and (c) the construction of an essentially self-adjoint generator $H$ of time evolution on $\mathcal H$. We can thus separate the quantization of the observable algebra, i.e. the construction of $(\mathfrak A,\mathcal H,\pi)$, from the time evolution, i.e. the construction of $H$. Once the first part is preformed, one can formally perform the second step by writing the domain of $H$ as $\oplus_{E} \mathcal H_E$, where $\mathcal H_E$ denotes the eigenspace of $H$ for eigenvalue $E$. 

The importance of the first step can not be underestimated, particularly if the theory is time-reparametrization invariant because in this case $H(p,q)$ is replaced with an energy conservation constraint $H(p,q)-E$. The second step becomes trivial in this case since the first step already implements the constraint $H(p,q)-E$. The physical Hilbert space $\mathcal H$ can thus be identified with $\mathcal H_E$. This simple situation is however complicated by the fact that at least classically one requires that the parametrization of trajectories is affine, which implies that the Lagrange-multiplier $N$ for the time-reparametrization constraint $H(p,q)-E$ is constrained to be positive. The implied linear ordering of classical trajectories is however lost in Dirac quantization, because the gauge orbits of the constraint $H(p,q)-E$ lack the ordering information that is available in the classical theory. 

This suggests that the standard Dirac quantization program should be supplemented such that the ordering information can be recovered, such as e.g. the thermal time hypothesis\cite{Connes:1994hv}. A natural proposal for sufficient structure to recover time ordering presents itself in the equivalence of gauge theory framework. The basic idea is to retain the ordering information by considering an ordered family of equivalent gauge theories, i.e. in the case of ADM gravity an ordered family of Shape Dynamics theories. To do this, we change the notion of equivalence of gauge theories from the previously used notion of dynamical equivalence to {\bf observable equivalence}, which we define as a particular coincidence condition for Poisson-algebras of Dirac-observables. This notion allows us to trade\footnote{Notice that the previously used notion of dynamical equivalence of gauge theories required that the time-reparametrization constraint was not traded, so time ordering is not a separate problem, since one can impose $N>0$ on both sides of the equivalence.} the time-reparametrization constraint $H(p,q)-E$ for a one-parameter family of constraints $Q(p,q)-\tau\approx 0$, where we require $\{Q(p,q),H(p,q)\}>0$, so the positivity of the lapse $N$ induces a $\frac{\partial \tau}{\partial t}>0$ through Hamilton's equations, which provides a linear order for the dual theories in terms of $\tau$. 

This suggests to Dirac quantize the one-parameter family of dual theories and define the quantization of the original system through the observable equivalence as precisely the same theory. Each dual theory, denoted by a value of $\tau$, is observationally equivalent to the original time-reparametrization-invariant theory at a particular instant of time. This gives a simple interpretation of the Dirac observables of the original theory as those observables of the dual theory that weakly coincide with the original observables at this instant of time. This suggests an experimental interpretation of the Dirac observables, which is an aspect of the global problem of time. Having an experimental interpretation of the Dirac observables makes it in principle, however not necessarily in practice, unnecessary to have a quantization of $H(p,q)$, because $H(p,q)$ is just a connection between the different interpretations of the observables of the original gauge theory. This is of course not a new point of view; very similar ideas are used to interpret timeless theories. The new development here is to use these ideas in the contect of equivalent gauge theories. This framework turns out to provide a natural and suggestive setting for these ideas.

Using this approach for Shape Dynamics addresses two problems: First, one evades the global non-local Hamiltonian and rather obtains a family of conformal theories that is dual to General Relativity. Second, one has a preferred interpretation of Dirac-observables in gravity, analogous to the notion introduced in sections 4.4.1 and 4.4.2 of \cite{HenneauxTeitelboim}. These are identified with Shape Dynamics observables at fixed York time and thus experimentally accessible by determining the conformal geometry of the universe at given York-time $\tau$.

This paper is organized as follows: We will provide some preliminaries on gauge theories and the precise relation between dynamical equivalence of gauge theories with the newly defined observable equivalence of gauge theories in section \ref{sec:ObservableEquivalence}. We will then consider the simplest possible toy model to motivate the approach outlined in this introduction in section \ref{sec:ToyModel} and then proceed with the construction of a one parameter family of ``Fixed Shape Dynamics'' theories in section \ref{sec:SDEquivalence}. To provide a concrete quantum gravity application, we apply this framework to Shape Dynamics in 2+1 dimensions in section \ref{sec:2Plus1} and briefly discuss the necessary steps to apply this procedure to 3+1 dimensions before we conclude in section \ref{sec:Conclusions}. The appendix provides a construction of the standard Shape Dynamics Hamiltonian that is useful for present paper using only standard arguments.

\section{Observable Equivalence}\label{sec:ObservableEquivalence}

In \cite{Gomes:2011zi} we defined two gauge theories A and B to be dynamically equivalent, if there is a distinct gauge for gauge theory A and a distinct gauge for gauge theory B, such that the trajectories of both theories coincide. This definition captures classical equivalence, but it is not very well suited for quantization, since trajectories arise only in a semiclassical limit. We thus argued in the introduction that one should define a notion of equivalence that puts observables (or perennials in Kucha\v{r}'s notation) to the forefront, since classical observables are translated into quantum operators, while classical trajectories lack a simple quantum analogue. 

\subsection{Preliminaries}

Let us start with recalling the textbook treatment of canonical gauge theories, in particular the description of observables, and the notion of linking gauge theories to make this paper reasonably self-contained. For this we assume that all second class constraints have been removed through phase space reduction, so we consider a canonical gauge theory $A=(\Gamma,\{.,.\},\{\chi_\alpha\}_{\alpha \in \mathcal A})$ where $\Gamma$ denotes a phase space that carries the Poisson-structure $\{.,.\}$ and where $\{\chi_\alpha\}_{\alpha \in \mathcal A}$ denotes a set of regular and irreducible first class constraints, which are imposed as initial value constraints. We assume that the theory is time-reparametrization invariant, so the dynamics is encoded in the energy conservations constraint $\chi_{\alpha=0}=H(p,q)-E$, where $q_i,p^j$ denote local Darboux-coordinates for $\Gamma$ with only nonvanishing elementary Poisson bracket $\{q_i,p^j\}=\delta^j_i$. The total Hamiltonian is $H_{tot}=\sum_{\alpha \in \mathcal A} \lambda^\alpha \chi_\alpha(p,q)$, where the $\lambda^\alpha$ are arbitrarily specifiable Lagrange multipliers, with the exception of $\lambda^{\alpha=0}=N$, which is constrained to be positive at all times. It follows form the first class property of the constraints that the dynamics generated by $H_{tot}$ preserves the constraint surface $\mathcal C:=\{(p,q)\in \Gamma: \chi_\alpha(p,q)=0 \forall \alpha\in\mathcal A\}$, so the physical dynamics is confined to $\mathcal C$ since physical initial conditions are in $\mathcal C$.

Measurements on classical systems are modeled by phase space functions\footnote{Throughout this paper we will always assume that the phase space functions satisfy suitable smoothness conditions so the Poisson structure is well defined.} $M(p,q)$ and observables $O(p,q)$ are phase space functions whose value at any later time is uniquely determined by the initial condition $(p,q)\in \mathcal C$. This implies the gauge-invariance condition $\left.\{O(p,q),\chi_\alpha(p,q)\}\right|_{\mathcal C}=0$. This description of observables is however redundant, since physical states of the theory are confined to the constraint surface. Independent observables are phase space functions that are gauge-invariant functions on $\mathcal C$, meaning that observables are (weak) equivalence classes of gauge invariant phase space functions, where weak equivalence means that the representatives coincide on $\mathcal C$. There are three natural characterizations of observables $\mathcal O(A)$ of a gauge theory $A$ and their Poisson algebra $(\mathcal O(A),\{.,.\}_A)$:

(1) {\it On phase space} $\Gamma$: An observable is an equivalence class of gauge-invariant phase space functions $[O(p,q)]_\sim$, where $O_1\sim O_2$ iff $\left.O_1\right|_{\mathcal C}=\left.O_2\right|_{\mathcal C}$, so observables are elements of the quotient of the functions on $\Gamma$ by the ideal $\mathcal I$ of functions that vanish on $\mathcal C$. Regularity of the constraint functions ensures that any element of $\mathcal I$ can be written as $\sum_{\alpha \in \mathcal A}f^\alpha(p,q)\chi_\alpha(p,q)$. It follows from the Jacobi-identity that the Poisson-bracket of any two representatives $f,g$ for observables $[f],[g]$ weakly commutes with all constraints, which allows one to define the Poisson structure for observables as $\{[f],[g]\}_A:=[\{f,g\}]_A$. This is well defined, i.e. independent of the representatives $f$ and $g$, because for $i_1,i_2 \in \mathcal I$ one has $[\{f+i_1,g+i_2\}]_A=[\{f,g\}]_A$. We denote the observables of $A$ together with the Poisson structure $\{.,.\}_A$ as the observable algebra $P(A)$. This description of the observable algebra in terms of representatives of equivalence classes of gauge-invariant functions is cumbersome, but in the presence of complicated constraints the only practically feasible description and in the case of field theories often the only local description.

(2) {\it On the constraint surface} $\mathcal C$: Observables can be equivalently described as gauge-invariant functions on the constraint surface $\mathcal C$, since we always quotient by $\mathcal I$. The Poisson structure in this description can be defined as $\{i^* f,i^* g\}_A:=\left.\{f,g\}\right|_{\mathcal C}$, where $i$ denotes the canonical embedding of the constraint surface into $\Gamma$ and $f,g$ denote functions on $\Gamma$. This is independent of the representatives $f,g$ due to the same argument as before and we can naturally identify the structure with the previously defined observable algebra $\mathcal P(A)$.

The first class property of the constraints provides an integrability condition when restricted to the constraint surface, providing that the constraint surface is foliated by gauge orbits. Let us introduce a set of regular irreducible gauge-fixing conditions\footnote{There is a high degree of symmetry between constraints and gauge fixing conditions, but we caution that the set of gauge fixing conditions is {\bf not} implied to be first class.} $\{\sigma^\alpha\}_{\alpha\in \mathcal A}$ that define a gauge fixing surface $\mathcal G:=\{(p,q)\in \Gamma:\sigma^\alpha(p,q)=0\forall \alpha \in \mathcal A\}$, such that each gauge orbit intersects $\mathcal G$ once and only once and $\{\sigma^\alpha,\chi_\beta\}$ is invertible on the reduced phase space $\Gamma_{red}=\mathcal C \cap \mathcal G$. 

(3) {\it On reduced phase space} $\Gamma_{red}$: This description of observables rests on the fact that gauge-invariant functions on $\mathcal C$ are completely determined by their restriction to $\Gamma_{red}$, since gauge-invariance implies constancy along gauge orbits. The Poisson-structure on $\Gamma_{red}$ can be described through the Dirac bracket $\{f,g\}_D:={f,g}-(\{f,\chi_\alpha\}(C^{-1})^\alpha_\beta\{\sigma^\beta,g\}-{f,\sigma^\beta}(C^{-1})^\alpha_\beta\{\chi_\alpha,g\})$, which weakly coincides with $\{.,.\}_A$ and has the important property that $\{f|_{\Gamma_{red}},g|_{\Gamma_{red}}\}_D=\{f,g\}|_{\Gamma_{red}}$, so it defines the desired Poisson structure for observables when these are expressed as functions on $\Gamma_{red}$. We can again identify the structure with the observable algebra $\mathcal P(A)$. Locally, one can always find adapted\footnote{Notice that the vanishing surfaces of the adapted coordinates do in general {\bf not} locally coincide with the constraint- and gauge fixing surfaces, only the intersection $\Gamma_{red}$ does locally coincide with the surface $p\equiv 0\equiv q$.} Darboux-coordinates $(p^\alpha,p_o^j;q_\beta,q^o_i)$, such that $\Gamma_{red}$ can be described as the surface $p\equiv 0\equiv q$, so the Dirac-bracket is defined through the only nonvanishing elementary bracket $\{q^o_i,p_o^j\}_D=\delta^j_i$.

Let us now turn to linking theories. A general linking theory $(A,\Sigma_1,\Sigma_2)$ is a gauge theory $A=(\Gamma,\{.,.\},\{\chi_\alpha\}_{\alpha\in \mathcal A})$ together with two sets of (partial) gauge fixing conditions $\Sigma_2$ and $\Sigma_2$. Let us now suppose that we can write $\Gamma=\Gamma_o\times\Gamma_e$ as the product of two phase spaces that Poisson-commute with each other and assume that the constraints can be split into three disjoint sets
\begin{equation}
 \mathcal X_1=\{\phi_\alpha-\chi^1_\alpha\}_{\alpha \in \mathcal A_o},\,\, \mathcal X_2=\{\pi^\alpha-\chi_2^\alpha\}_{\alpha \in \mathcal A_o}, \mathcal X_3=\{\chi^3_\mu\}_{\mu \in \mathcal M},
\end{equation}
where the time-reparametrization constraint is contained in $\mathcal X_3$ and where $(\phi_\alpha,\pi^\beta)$ is a canonically conjugate pair furnishing Darboux-coordinates for $\Gamma_e$ and where the $\chi^1_\alpha,\chi_2^\beta,\chi^3_\mu$ are functions on $\Gamma_o$, so one can then impose the special gauge-fixing conditions 
\begin{equation}
  \Sigma_1=\{\phi_\alpha\}_{\alpha \in \mathcal A}, \,\,\,\Sigma_2=\{\pi^\alpha\}_{\alpha \in \mathcal A}. 
\end{equation}
One then finds that the reduced phase space is $\Gamma_o$  for both phase space reductions and that both Dirac-brackets coincide with the Poisson-bracket $\{.,.\}_o$ on $\Gamma_o$. One can now impose the gauge fixing conditions $\chi^1_\alpha=0$ and $\chi_2^\beta=0$ and impose $N=1$ to show that the two gauge theories $B=(\Gamma_o,\{.,.\}_o,\mathcal X_1\cup\mathcal X_3)$ and $C=(\Gamma_o,\{.,.\}_o,\mathcal X_2\cup\mathcal X_3)$ have coinciding gauge-fixed trajectories and are thus dynamically equivalent. Let us conclude with an almost trivial but important remark: The notion of dynamical equivalence requires that the time-reparametrization constraint is contained in the set $\mathcal X_3$ of constraints that was not traded by going from $B$ to $C$. 

\subsection{Observable Equivalence}

Let us now develop a notion of equivalence of gauge theories in terms of Dirac observables. We start by spelling out the physical requirements that we want to retain for the purpose of quantization. The first step of quantization of a canonical gauge theory $A$ is the construction of an associative observable algebra $O(A)$, that quantizes the classical Poisson algebra $\mathcal P(A)$. Quantization means that one has a map $q:\mathcal P(A) \to O(A)$ such that $q(f)q(g)=q(fg)+\mathcal O(\hbar)$ and $[q(f),q(g)]=i\hbar\left(q(\{f,g\})+\mathcal O(\hbar)\right)$, where these requirements are usually replaced by the stronger condition that for a Poisson-closed set of generators $\mathcal G(A)$ of $\mathcal P(A)$ one imposes $[q(f),q(g)]=i\hbar q(\{f,g\})$ for all $f,g\in \mathcal G(A)$. The problem of quantization of gauge theories is to find a quantization map for the observable algebra. It is often particularly useful to follow the Dirac procedure and construct the quantization of the observable algebra in a two-step process.

If one is able to find a classical observable algebra $\mathcal P(A)$ that is Poisson-isomorphic to an observable algebra $\mathcal P(B)$, i.e. there is a Poisson-morphism $i:\mathcal P(A) \to \mathcal P(B)$ and if one has a quantization map $q:\mathcal P(B) \to O(B)$ then $q\circ i$ is a quantization of $\mathcal P(A)$, because the two quantization conditions $q(i(f))q(i(g))=q(i(fg))+\mathcal O(\hbar)$ and $[q(i(f)),q(i(g))]=i\hbar\left(q(i(\{f,g\}))+\mathcal O(\hbar)\right)$ are satisfied because if they are satisfied for $\mathcal P(B)$ because $i$ is a Poisson-morphism. We want this to be implied by observable equivalence, since it is very useful to be able to reinterpret the quantization of one theory as the quantization of another. We thus say
``We call two gauge theories $A$ and $B$ canonically observable equivalent, iff there is a gauge theory $C$ and two Poisson isomorphisms $i_A: \mathcal P(A) \to \mathcal P(C)$ and $i_B: \mathcal P(B) \to \mathcal P(C)$ between the observable algebras $\mathcal P(A)$, $\mathcal P(B)$ and $\mathcal P(C)$ of $A$,$B$ and $C$.''

The concatenations $i_A^{-1} \circ i_B$ and $i_B^{-1} \circ i_A$ are then Poisson morphisms between $\mathcal P(A)$ and $\mathcal P(B)$ that allow us to interpret quantizations of theory $A$ as a quantization of theory $B$ and vice versa. We will refine this definition in the next subsection to ``observable equivalence'' that has additional structure coming form a linking theory.

Let us conclude this definition with emphasizing that the definition of canonical observable equivalence does allow for the time-reparametrization constraint is traded, which was the source of technical difficulties in the general formalism and in particular the source of the nonlocality of the Shape Dynamics Hamiltonian. However it is possible to trade the time-reparametrization constraint in the case of canonical observable equivalence due to the fact that Dirac-observables are required to weakly commute with this constraint, thus giving a timeless picture of Dirac observables (or perennials). We will address the structure that needed for recovery of dynamics form the timeless picture in section \ref{sec:GeneralRecovery}.

\subsection{Observable Equivalence from Linking Theories}

Linking theories are a very useful tool to prove dynamical equivalence of gauge theories. Let us now derive (canonical) observable equivalence form linking theories, to have this  tool at our disposal for establishing (canonical) observable equivalence. For this we assume that we have a special linking theory with gauge theory $A=(\Gamma_o\times\Gamma_e,\{.,.\},\{\phi_\alpha-\chi^1_\alpha\}_{\alpha \in \mathcal A},\{\pi^\alpha-\chi_2^\alpha\}_{\alpha\in\mathcal A},\{\chi^3_\mu\}_{\mu \in \mathcal M})$ as described above and the two sets of partial gauge fixing conditions $\Sigma_1=\{\phi_\alpha\}_{\alpha \in \mathcal A}$ and $\Sigma_2=\{\pi^\alpha\}_{\alpha\in\mathcal A}$. 

We saw that if the gauge fixing condition is proper, i.e. each gauge orbit is intersected once and only once by the gauge fixing surface, then we can identify observables with functions of a suitable smoothness class on reduced phase space. To use this identification, we consider the dictionary theory $(\Gamma_{red},\{.,.\}_D,\{\chi^3_\mu\}_{\mu\in\mathcal M})$, where $\Gamma_{red}=\{x\in \Gamma_o:\chi^1_\alpha(x)=0=\chi_2^\alpha(x)\forall \alpha \in \mathcal A\}$ and $\{.,.\}_D$ denotes the Dirac bracket associated with the phase space reduction $\Gamma_o\to\Gamma_{red}$. Notice that the $\chi_2$-constraints are by construction a gauge fixing for the $\chi_1$-constraints and vice versa if the Dirac matrix $\{\chi_1^\alpha,\chi^2_\beta\}$ is invertible, but to make the identification of observables with unrestricted smooth phase space functions on $\Gamma_{red}$, we need to assume that this gauge fixing is proper. Let $r: C^{a}(\Gamma_o) \to C^{b}(\Gamma_{red}$ denote the restriction map, i.e. $r(f):=\left. f\right|_{\Gamma_{red}}$, where $a$ and $b$ denote smoothness classes that are compatible with the restriction $\Gamma_{o} \to \Gamma_{red}$. Furthermore, we notice that the restriction of an $A$-observable is independent of its particular representative, i.e. if $O_1$ and $O_2$ are two representative functions on $\Gamma_o$ of the observable $[O]_A$, then $r(O_1)-r(O_2)=r(O_1-O_2)=0$, where we used linearity of the restriction map and that representative functions coincide on the constraint surface. The analogous statement holds for $B$-observables, so we have the two maps
\begin{equation}
 \begin{array}{rcl}
  i_A([O]_A) & := & r([O]_A)\\
  i_B([O]_B) & := & r([O]_B), 
 \end{array}
\end{equation}
which are defined through the restriction of an arbitrary representative function and differ only through the domains that they act on. To establish the Poisson-isomorphism property we need the to use the Poisson-morphism property of the Dirac bracket, i.e. $r(\{[O_1]_A,[O_2]_A\})=\{r([O_1]_A),r([O_2]_)\}_D$, and the bijectivity of $i_A$ and $i_B$, which follows from the domains of the two maps. We can thus combine $i_{BA}=i_A^{-1}\circ i_B$ and $i_{AB}=i_B^{-1}\circ i_A$ to identify the two observable algebras with each other. This establishes what we call canonical observable equivalence.

Let us now consider the structure that comes from the linking theory: First, the two equivalent gauge theories $A$ and $B$ are defined on the same phase space with the same Poisson structure $(\Gamma,\{.,.\})$. Second, the phase space $\Gamma_{red}$ of the dictionary theory $C$ is embedded into $\Gamma$ by an embedding $i$ whose pull-back extends to the Poisson-isomorphisms between the observable algebras. This additional structure is useful, so we define
\begin{defi} 
  We call two gauge theories $A=(\Gamma,\{.,.\},\{\chi^1_\alpha\}_{\alpha \in \mathcal A})$ and $B=(\Gamma,\{.,.\},\{\chi_2^\beta\}_{\beta \in \mathcal B})$ on the same pahse space $(\Gamma,\{.,.\})$ {\bf observable equivalent} if there is a gauge theory $C=(\Gamma_{red},\{.,.\}_D,\{\chi_3^\mu\}_{\mu \in \mathcal M})$ and an embedding $i:\Gamma_{red} \to \Gamma$ whose pull-back extends to a Poisson-isomorphism between the observable algebras $\mathcal P(A),\mathcal P(B)$ and $\mathcal P(C)$.
\end{defi} 
Using this definition we have:
\begin{prop}
 If there is a linking gauge theory that links gauge theory $A$ and $B$ then $A$ and $B$ are observable equivalent.
\end{prop}
This statement is not new, it is just text-book knowledge expressed in the context of observable equivalence of gauge theories. The value of this reformulation is that is motivated by using observable equivalence as a tool for quantizing of gauge theories.

\subsection{Adapted Phase Space}

A particularly useful application of observable equivalence of gauge theories is the case where theory $A=(\Gamma,\{.,.\},\{\chi^1_\alpha\}_{\alpha \in \mathcal A})$ has complicated observables, but an observable equivalent theory $B=(\Gamma,\{.,.\},\{\chi_2^\alpha\}_{\alpha \in \mathcal A})$ in the same phase space $(\Gamma,\{.,.\})$ allows us to construct an adapted phase space, i.e. we can satisfy the following definition:
\begin{defi}
  Given a gauge theory $B=(\Gamma_o,\{.,.\},\{\chi^2_\alpha\}_{\alpha\in \mathcal A})$, we $(\Gamma_1,\Gamma_2)$ {\bf adapted} to $B$ if $\Gamma_o=\Gamma_1\times\Gamma_2$ and
  \begin{enumerate}
   \item any smooth function on $\Gamma_1$ Poisson commutes with any smooth function on $\Gamma_2$,
   \item the constraints $\{\chi_2^\alpha=p_2^\alpha\}_{\alpha \in \mathcal A}$ are on $\Gamma_2$ and
   \item the gauge orbits of are the sets $S(x)=\{(x,y)\in \Gamma_1\times\Gamma_2:\chi^\alpha_2(y)=0\forall \alpha\in \mathcal A\}$,
  \end{enumerate}
\end{defi}
where $(q^1_i,p^j_1)$ denotes local Darboux coordinates for $\Gamma_1$ and $(q^2_\alpha,p_2^\beta)$ denotes Darboux coordinates for $\Gamma_2$. The implicit function theorem provides that the constraints of theory $A$ can be (locally) solved for $q$, i.e. that they can be replaced with the equivalent set of constraints  $\{\chi^1_\alpha=q^2_\alpha-\xi_\alpha(q^1,p_1)\}_{\alpha\in\mathcal A}$, which can be shown to be Abelian by a standard argument. 

Using adapted coordinates it is straightforward to construct representative functions for $B$-observables, since all function of the form $f_{B}(p_1,q^1;p_2)\approx f_{B}(p_1,q^2;p_2)$ automatically commute with all $\chi_2^\alpha$. The vector fields $L_\alpha: f \mapsto \{\chi^1_\alpha,f\}$ satisfy
\begin{equation}
 [L_\alpha,L_\beta]: f \mapsto \{\chi^1_\alpha,\{\chi^1_\beta,f\}\}-\{\chi^1_\beta,\{\chi_\alpha,f\}\}=\{f,\{\chi^1_\alpha,\chi^1_\beta\}\}=0
\end{equation}
since the $\chi^1_\alpha$ are Abelian. The Frobenius theorem thus provides local integrability, so we can in principle find a complete set of local phase space functions $Q_i(p_1,q^1;p_2)$ and $P^j(p_1,q^1;p_2)$ that satisfy 
\begin{equation}
 \begin{array}{ll}
   Q_i(p_1,q^1;0)=q^1_i,&0=\{q^2_\alpha-\xi_\alpha,Q_i\}=\frac{\partial Q_i}{\partial p_2^\alpha}-\{\xi_\alpha,Q_i\},\\
   P^i(p_1,q^1;0)=p_1^i,&0=\{q^2_\alpha-\xi_\alpha,P^i\}=\frac{\partial P^i}{\partial p_2^\alpha}-\{\xi_\alpha,P^i\}.
 \end{array}
\end{equation}
One can thus in principle find local representative functions for $A$-observables, because any local phase space function of the form $f_{A}(P,Q;q_1-\xi)\approx f_{A}(P,Q;0)$ Poisson commutes with all $q^2_\alpha-\xi_\alpha$.

Let us now restrict the observables to the intersection $\Gamma_{red}$ of the constraint surfaces $\mathcal C_A$ and $\mathcal C_B$ of $A$ and $B$, i.e. $p_2^\alpha=0=q^2_\alpha-\xi_\alpha$:
$f_{B}(q_1,p^1;p_2)|_{\mathcal C_A}=f_B(q_1,p^1;0)$ and $f_A(Q,P;q^2-\xi)|_{\mathcal C_A\cap \mathcal C_B}=f_A(Q,P;0)|_{\mathcal C_B}=f_A(q,p;0)$. We thus have a very simple identification of the an $A$ observable $f_A$ with a $B$ observable $f_B$ if
\begin{equation}
  f_A(q_1,p^1;0)=f_B(q_1,p^1;0)\,\forall (q_1,p^1)\in \Gamma_1.
\end{equation}
The practical value of this identification is that it does not explicitly refer to the $(Q_i,P^j)$, which are very hard to construct, but only needs the much simpler $(q_1,p^1)$ and the notion of $\Gamma_{red}$. Moreover, the expression for the Dirac-bracket on $\Gamma_{red}$ simplifies to the Poisson bracket for functions that just depend on $\Gamma_2$:
\begin{equation}
 \{f,g\}_D=\{f,g\}-(\{f,\chi_\alpha^1\}{C^{-1}}^\alpha_\beta\{\chi_2^\beta,g\}-\{f,\chi^\beta_2\}{C^{-1}}^\alpha_\beta\{\chi_1^\alpha,g\})=\{f,g\},
\end{equation}
where $C$ denotes the Dirac matrix. This is a significant simplification for observables, because it follows straightforwardly that observables can be identified with equivalence classes of phase space functions on $\Gamma_1$ that weakly Poisson commute with the constraints $\{\chi^3_\mu\}_{\mu\in \mathcal M}$, where equivalence means weakly coinciding.

This identification of the Dirac algebra of $B$-observables with the Poisson algebra on $\Gamma_2$ is also nothing new; it was already used by Dirac as a motivation to introduce the Dirac bracket. However it has a nontrivial consequence in the context of observable equivalence: If $A$ and $B$ are observable equivalent, then we can use the Poisson-isomorphism $i_{BA}$ to identify the Poisson algebra on $\Gamma_2$ with the Dirac algebra of $A$-observables. This leads to the important simplification that we do not have to construct the Dirac algebra of $A$-observables! In other words: We can {\bf reinterpret} the elements of the very simple algebra of $B$-observables as a particular phase space functions, which when evaluated in $B$-ague coincide with unique representatives of the observable algebra of theory $A$.

\section{Toy Model and Recovery of Dynamics}\label{sec:ToyModel}

We replaced dynamical equivalence with observable equivalence to be able to trade the time-reparametrization constraint, while maintaining sufficient structure be able to claim that a quantization of an observable equivalent theory can be interpreted as a quantization of the original theory. However by trading the time-reparametrization constraint, we loose the classical time-ordering information. We will now consider the simplest nontrivial toy model to gain insight into the structure that is necessary to recover this information and extract a proposal for recovering dynamics in the general case.

\subsection{Toy Model Setup}

The simplest nontrivial example is a free particle with energy $E$ in two dimensions with the simplifying assumption $p_2\ne 0$, so the degenerate case $E=0$ is excluded. We have configuration variables $(q_1,q_2)$ and canonically conjugate momenta $(p^1,p^2)$ and a time-reparametrization (or energy conservation) constraint $\chi=\vec{p}^2-E$. The total Hamiltonian of this original theory is $ H^{orig}_{tot}=N \left(\frac 1 2 \vec{p}^2-E\right) \approx 0$, where forward evolution in time requires $N>0$. We will refer to this theory as $A$.

$A$ arises form the linking theory with constraints $\tilde\chi=\frac 1 2 (p_1^2+(p_2+\pi)^2)\approx 0$ and $\tilde \sigma=q_2-\phi-\tau \approx 0$, where $\phi,\pi$ is an auxiliary canonical pair, through the imposition of the partial gauge fixing $\pi=0$, which yields the phase space reduction $(\phi,\pi)\to(q_2-\tau,0)$ and theory $A$. Imposing the partial gauge fixing condition $\phi=0$ yields the phase space reduction $(\phi,\pi)\to(0,\sqrt{2\epsilon-p_1^2}-p_2)$, which reveals the equivalent gauge theory constraint $\sigma=q_2-\tau \approx 0$ and total dual Hamiltonian $ H^{dual}_{tot}=\rho\left(q_2-\tau\right)$; we will refer to this theory as $B$.

The B-adapted phase space is trivial, we write $\Gamma=\Gamma_1(q_1,p_1)\times\Gamma_2(q_2,p_2)$, so the representative functions for $B$-observables are given by
\begin{equation}
 f_B(q_1,p_1;\sigma)\,\approx_B\, f_B(q_1,q_2;0).
\end{equation}
The simplicity of the system allows us to find variables $(Q,P)=(q_1-\frac{p_1}{p_2}(q_2-\tau),p_1)$ that Poisson commute with $\chi$ and coincide with $(q_1,p_1)$ on the surface $\sigma=0$, so representatives for $A$-observables are
\begin{equation}
 f_A(Q,P;\chi)\,\approx_A\,f_A(Q,P;0).
\end{equation}
Evaluating these phase space functions on the intersection of the surface $\chi=0$ with the surface $\sigma=0$ gives $f_A(Q,P;0)|_{\sigma=0}=f_A(q_1,p_1;0)$, as expected. This gives the identification
\begin{equation}\label{equ:ToyIdentification}
 f_A \leftrightarrow f_B \textrm{ iff } f_A(q_1,p_1;0)|_{\sigma=0}=f_B(q_1,p_1;0)|_{\chi=0}\,\forall (q_1,p_1) \in \Gamma_1.
\end{equation}
This is analogous to the interpretation in section 4.4 of \cite{HenneauxTeitelboim}.

\subsection{Recovering Dynamics in Toy Model}\label{sec:GeneralRecovery}

We know the dynamics of a free particle at fixed energy and how to reconstruct its trajectory from the timeless data $(Q,P)$ and an orientation; it is the purpose of this subsection to view this reconstruction in the context of observable equivalence. The trajectory $T(\vec q_o,\vec p_o)$ as a set of points (i.e. without direction) is the one-diemsnional subset of points $T(\vec q_o,\vec p_o)=\{Q=q_o,P=P_o\}$ in phase space. The direction of time along the trajectory is given by the vector field $v_\chi=f\mapsto \{\chi,f\}$.  

It is clear that we can not recover this directional information form representative functions of Dirac-observables of theory $A$, since these are invariant under the flow generated by $v_\chi$. Representative functions $f_B(q_1,p_1;\sigma)$ of observables of theory $B$ are however not invariant under this flow, but using the equivalent constraint $\tilde \chi=p_2-\xi\approx 0$, where $\xi=\sqrt{2E-p_1^2}$ due to the assumption $p_2>0$, we have
\begin{equation}
 \begin{array}{rcl}
    \left.\{\tilde \chi,f_B(q_1,p_1;q_2)\}\right|_{\sigma=0}&=&-\left.\frac{\partial f_B(q_1,p_1;q_2)}{\partial q_2}\right|_{q_2=\tau}-\left.\{\xi,f_B(q_1,p_1;q_2)\}\right|_{q_2=\tau}\\
    &=&-\frac{\partial f_B(q_1,p_1;\tau)}{\partial \tau}-\left.\{\xi,f_B(q_1,p_1;q_2)\}\right|_{q_2=\tau},
 \end{array}
\end{equation}
which vanishes only for particular representatives of $B$-observables that would be called ``perennials'' by Kucha\v{r}. This form suggests not to consider one particular dual theory $B$, but a family of dual theories $B(\tau)$ and to recover the time-ordering of theory $A$ from an ordering of this family of dual theories with constraints $\sigma=\sigma_o-\tau$. This ordering is induced by the evolution of $\sigma_o$ through Hamilton's equations $\dot \sigma_o=N\{\sigma_o,\chi\}=N p_2>0$ in theory $A$, since we assumed $p_2>0$. We thus recover from the preservation of the constraint $d\sigma=0$ the direction of the trajectories as $d\tau>0$. 

Moreover, one can recover the point sets of a trajectory of theory $A$ from Dirac observables as the set of points in phase space that is fixed by fixing a maximal number of independent Dirac observables in theory $A$. This together with observable equivalence of theory $A$ with theory $B(\tau)$ at any value of $\tau$ suggests to recover the dynamics of theory $A$ in a two-step process: First identifying observables through equivalence and then recovering the time-ordering of trajectories of $A$ from the $\tau$-ordering of dual theories. 

The toy model thus leads to the proposal to consider a gauge theory $A=(\Gamma,\{.,.\},\{\chi_\alpha\}_{\alpha\in \mathcal A})$ with a time-reparametrization constraint $\chi_o$ that is observable equivalent to a one-parameter family of gauge theories $B(\tau)=(\Gamma,\{.,.\},\{\sigma_\alpha\}_{\alpha\in \mathcal A})$ with constraint $\sigma_o-\tau$ for which  the strict inequality $\{\chi_o,\sigma_o\}>0$ (or strictly $\{\chi_o,\sigma_o\}<0$) as well as $\{\chi_{\alpha\ne o},\sigma_o\}\approx 0$ holds. Notice that this dynamical equivalence is infinitesimal as does not give matching trajectories, but only  matching initial conditions and a matching direction of time that is recovered from $d\tau>0$ (or $d\tau <0$). To recover the complete trajectories beyond the infinitesimal level one has to replace the direction $d\tau$ with the vector field generated by $\tilde \chi$ in the toy model, which is as expected equivalent with not trading the time-reparametrization constraint.

From a purely formal perspective, the one parameter family of dual theories is only a particular way to label which of the original constraints generates dynamics and thus requires a strictly positive Lagrange-multiplier. There is however a straightforward physical interpretation: The requirements $\{\chi_o,\sigma_o\}>0$ and $\{\chi_{\alpha\ne o},\sigma_o\}\approx 0$ imply that $\sigma_o$ is a good clock variable for theory $A$, that evolves strictly monotonically. The constraints $\sigma_o-\tau$ indicate the instant in time w.r.t. the $\sigma_o$-clock and the ordering w.r.t. increasing $\tau$ is the ordering induced by the time on this clock.

\section{Observable Equivalence between General Relativity and Shape Dynamics}\label{sec:SDEquivalence}

Let us now construct a one-parameter family of Shape Dynamics; we will refer to the members of this family as ``Fixed Shape Dynamics'' to distinguish it from the version of Shape Dynamics with generator of dynamics. General Relativity and Fixed Shape Dynamics are both formulated as gauge theories on ADM-phase space. For this we fix a compact Cauchy surface $\Sigma$ without boundary and use the spatial metric $g_{ab}(x)$ and its conjugate momentum density $\pi^{ab}(x)$ with elementary Poisson bracket
\begin{equation}
 \{g_{ab}(x),\pi^{cd}(y)\}=\delta^{(3)}(x,y) \delta^{cd}_{(ab)}.
\end{equation}
Both theories are defined to have an on-shell vanishing Hamiltonian and are assumed to be gauge theories of spatial diffeomorphisms, i.e. the set of constraints contains the spatial diffeomorphism constraints 
\begin{equation}
  H(v)=\int_\Sigma d^3x \pi^{ab}(\mathcal L_v g)_{ab}\approx 0,
\end{equation}
where $\mathcal L_v$ denotes the Lie-derivative w.r.t. an arbitrary vector-field $v$ on $\Sigma$.

\subsection{Linking Theory}

To construct Shape Dynamics we extend the ADM phase space $\Gamma_{ADM}$ by the phase space $\Gamma_\phi$ a scalar field $\phi$ and its canonically conjugate momentum density $\pi_\phi$. Dynamically equivalent Shape Dynamics is constructed form the following system with Lagrange multipliers $N$, $\rho$ and $v^a$:
\begin{equation}
 \begin{array}{rcl}
   H_{tot}&=&S(N)+H(v)+Q(\rho)\\
   S(N)&=&\int d^3x N\left(\frac{1}{\sqrt{|g|}}\pi^{ab}(g_{ac}g_{bd}-\frac 1 2 g_{ab}g_{cd})\pi^{cd}-R\sqrt{|g|}\right)\\
   H(v)&=&\int d^3x \pi^{ab}\mathcal L_v g_{ab} \\
   Q(\rho)&=&\int d^3x \rho \pi_\phi, 
 \end{array}
\end{equation}
by ``Kretschmannization'' w.r.t. volume-preserving conformal transformations, which is implemented through a canonical transformation generated by $F=\int d^3x(g_{ab}e^{4\hat \phi}\Pi^{ab}+\phi \Pi_\phi)$, where the operation $\phi \to \hat \phi$ implements volume preservation. To construct Fixed Shape Dynamics we are tempted to use a similar generating functional $ F=\int_\Sigma d^3x\left(g_{ab}e^{4\phi}\Pi^{ab}+\phi \Pi_\phi +\tau \sqrt{|g|}\right)$, which generates a promising canonical transformation $g_{ab}\to e^{4\phi} g_{ab}$, $\pi^{ab}\to e^{-4\phi}\left(\pi^{ab}-\frac\tau 2 \sqrt{|g|}g^{ab}\right)$, $\phi\to\phi$ and $\pi_\phi\to\pi_\phi-4\left(\pi-\frac 3 2 \tau \sqrt{|g|}\right)$.
However, a different strategy turns out to be more fruitful: We start with standard Shape Dynamics, which is dynamically equivalent to ADM-gravity, with total Hamiltonian
\begin{equation}
   H_{tot}=N H_{SD}+Q(\rho)+H(v),
\end{equation}
where $Q(\rho)=\int d^2x \rho (\pi-\langle \pi\rangle \sqrt{|g|})$ is the volume preserving conformal constraint. We then trade $H_{SD}$ for a one-parameter family of global conformal constraints $Q=\frac 2 3 \langle \pi \rangle - \tau$. To construct the linking theory we use that $H_{SD}$ is equivalent to $V-V_o[\rho,\sigma]$ (cf. appendix), where $V_o$ denotes the volume that is ``corrected'' by the conformal factor that solves the Lichnerowicz-York equation in terms of shape data and is thus conformaly invariant. To construct the linking theory, we extend phase space by a single degree of freedom $q$ and its canonically conjugate momentum $p$ and consider the system
\begin{equation}
  \begin{array}{rcl}
    H_{SD}&=&q-V-V_o\\
    Q&=&p+2/3 \langle\pi\rangle-\tau\\
    H(v)&=&\int d^3x \pi^{ab}(\mathcal L_vg)_{ab}\\
    Q(\rho)&=&\int d^3x \rho (\pi-\langle \pi\rangle \sqrt{|g|}),
 \end{array}
\end{equation}
the two sets of partial gauge fixing conditions $q=0$ to recover General Relativity and $p=0$ to construct Fixed Shape Dynamics.

\paragraph{Constructing Fixed Shape Dynamics}

The gauge fixing condition $p=0$ leads to the phase space reduction $(q,p)\to(V-V_o,0)$ which trivializes $H_{SD}$ and turns $Q\to 2/3 \langle\pi\rangle-\tau$. The parametrization of the constraints $Q(\rho)$ is redundant, since $Q(\rho+\alpha)=Q(\rho)$ for any spatial constant $\alpha$. It is thus convenient to combine $Q(\rho)+3/2 \langle\rho\rangle V Q$ to lift this redundancy in the parametrization. This phase space reduction yields Fixed Shape Dynamics on the ADM phase space (where the Dirac bracket coincides with the Poisson bracket):
\begin{equation}
 \begin{array}{rcl}
   H_{tot}&=&Q(\rho)+H(v)\\
   Q(\rho)&=&\int d^3x \rho \left(\pi-\frac 3 2 \tau \sqrt{|g|}\right)\\
   H(v)&=&\int d^3x \pi^{ab}\mathcal L_v g_{ab},
 \end{array}
\end{equation}
which is a gauge theory of spatial diffeomorphisms and conformal transformations.

\paragraph{Recovering General Relativity}

To recover General Relativity, we impose the gauge fixing condition $q=0$, which leads to the phase space reduction
\begin{equation}
 (q,p)\to(0,2/3 \langle\pi\rangle-\tau).
\end{equation}
This phase space reduction trivializes $Q$ gives usual Shape Dynamics, which is dynamically equivalent to ADM-gravity (on ADM phase space with the ADM-Poisson bracket). We have thus shown observable equivalence between Shape Dynamics at fixed $\tau$ and ADM-gravity. 

\subsection{Adapted Phase Space}

The observable equivalence between General Relativity and Fixed Shape Dynamics allows us to identify the Dirac observables of both theories. In particular, there is an intuitive interpretation of a Fixed Shape Dynamics observable $O$, as the outcome of a certain measurement on a conformal metric theory with absolute simultaneity.

It is useful to re-express this in adapted variables, which are defined using a reference density $\omega(x)$ on $\Sigma$
\begin{equation}
 \begin{array}{rcl}
   \rho_{ab}(x)&:=&|g(x)|^{-1/3} g_{ab}(x)\\
   \sigma^{ab}(x)&:=&|g(x)|^{1/3}(\pi^{ab}(x)-g^{ab}(x)\pi(x))\\
   \phi(x)&:=&\frac 1 6 \ln\left(\frac{\sqrt{|g|(x)}}{\omega(x)}\right)\\
   \pi(x)&:=&\pi(x),
 \end{array}
\end{equation}
hence $g_{ab}(x)=e^{4\phi(x)}\omega^{2/3}(x)\rho_{ab}(x)$ and $\pi^{ab}(x)=\omega^{-2/3}(x)e^{-4\phi(x)}\left(\sigma^{ab}(x)+\frac 1 3 \rho^{ab}(x)\pi(x)\right)$, so the conformal constraints can be written as
\begin{equation}
 Q(\rho)=\int d^3x \rho(x)\left(\pi(x)-\lambda \omega(x)e^{6\phi(x)}\right).
\end{equation}
This is an adapted phase space for the conformal part of Fixed Shape Dynamics, because the variables $(\phi,\pi)$ Poisson commute with $(\rho_{ab},\sigma^{ab})$, so $\Gamma_1$ is coordinatized by $(\rho_{ab},\sigma^{ab})$ and $\Gamma_2$ is coordinatized by $(\phi,\pi)$. Shape Dynamics observables can thus be represented by diffeomorphism-invariant functionals of $(\rho_{ab},\sigma^{ab})$. In light of our general discussion of adapted phase space, we interpret $O[\rho,\sigma]$ as a representative function for the Dirac-observable of General relativity $O[P,\Sigma;\langle\pi\rangle]$, where $(P_{ab},\Sigma^{ab})|_{\langle\pi\rangle=\tau}=(\rho_{ab},\sigma^{ab})$ and commute with $H_{SD}$, that coincides with $O[\rho,\sigma]$ at fixed York time $\tau$.

\section{Quantum Gravity}\label{sec:2Plus1}

\subsection{Revisiting 2+1}

The probably most useful nontrivial quantum gravity model is the torus universe in 2+1 dimensions, which we now consider to investigate the consequences of the proposed quantization strategy. 

\subsubsection{Classical Theory}

Let us use the result of \cite{Budd:2011er} where we followed \cite{Moncrief:1989dx,Carlip:book} and explicitly constructed Shape Dynamics on the torus and the dynamical equivalence with ADM was established. It is convenient to fix a global chart $(x^1,x2)\in[0,1)^2$ and consider the reference metrics and reference  momenta
\begin{equation}
 \begin{array}{rcl}
   \bar g&=&\frac 1{\tau_2}\left(dx^1\otimes dx^1+(\tau_1^2+\tau_2^2)dx^2\otimes dx^2+\tau_1(dx^1\otimes dx^2 +dx^2 \otimes dx^1)\right)\\
   p&=&\frac 1 2\left(((\tau_1^2+\tau_2^2)p_2-2\tau_1\tau_2 p^1)\partial_1\otimes \partial_2 +p^2 \partial_2\otimes \partial_2 +(\tau_2p^1-\tau_1p^2)(\partial_1\otimes \partial_2+\partial_2 \otimes \partial_1)\right)
 \end{array}
\end{equation}
and write an arbitrary metric $g_{ab}=e^{2\lambda}(f^*g)_{ab}$, where $\lambda$ denotes a conformal factor and $f$ a diffeomorphism, and an arbitrary momentum $\pi^{ab}=e^{-2\lambda}(p^{ab}+\frac 1 2 \pi g^{ab}+\sqrt{|\bar g|}CY^{ab})$, where $CY^{ab}$ denotes the conformal Killing form w.r.t. a vector field $Y$, so $\{\tau_i,p^j\}=\delta^j_i$. Using these definitions, we can write Shape Dynamics as
\begin{equation}
 \begin{array}{rcl}
   H_{tot}&=&N H_{SD}+H(v)+Q(\rho)\\
   H_{SD}&=&\frac{\tau_2^2}{2V}((p^1)^2+(p^2)^2)-\frac{V}{2}(\langle \pi \rangle^2-4\Lambda)\\
   H(v)&=&\int d^2x \pi^{ab}(\mathcal L_vg)_{ab}\\
   Q(\rho)&=&\int d^2x \rho (\pi-\langle \pi\rangle \sqrt{|g|}).
 \end{array}
\end{equation}
We can solve the Shape Dynamics Hamiltonian $H_{SD}$ for the volume to get the equivalent constraint
\begin{equation}
 V-\tau_2\sqrt{\frac{(p^1)^2+(p^2)^2}{\langle\pi\rangle^2-4\Lambda}}\approx 0
\end{equation}
where we use a tile of Teichm\"uller space with $\tau_2>0$ and used the physical condition $V>0$ to have a unique solution for $\langle \pi\rangle >2\sqrt{\Lambda}$. This form shows that we can trade the Shape Dynamics Hamiltonian $H_{SD}$ for a one-parameter family of constraints $Q_o=\langle\pi\rangle-\tau$. To construct the equivalence explicitly we proceed as in 3+1 dimensions and extend phase space by a single degree of freedom $q$ with canonically conjugate momentum $p$ and consider the linking theory with the following constraints:
\begin{equation}
  \begin{array}{rcl}
    H&=&q-V+\tau_2\sqrt{\frac{(p^1)^2+(p^2)^2}{\langle\pi\rangle^2-4\Lambda}}\\
    Q&=&p+\langle\pi\rangle-\tau\\
    H(v)&=&\int d^2x \pi^{ab}(\mathcal L_vg)_{ab}\\
    Q(\rho)&=&\int d^2x \rho (\pi-\langle \pi\rangle \sqrt{|g|}).
  \end{array}
\end{equation}
Imposing the gauge-fixing condition $q=0$ leads to the phase space reduction $(q,p)\to(0,\tau-\langle\pi\rangle)$ and reduces the system to Shape Dynamics. Imposing the gauge fixing condition $p=0$ leads to the phase space reduction $(q,p)\to(V-\tau_2\sqrt{\frac{(p^1)^2+(p^2)^2}{\langle\pi\rangle^2-4\Lambda}},0)$ and results in a theory with constraints $Q_o$, $H(v)$ and $Q(\rho)$. Combining $Q_o$ and $Q(\rho)$ as above gives the one-parameter family of observable equivalent theories:
\begin{equation}
 \begin{array}{rcl}
  H_{tot}&=&Q_\tau(\rho)+H(v)\\
  Q_\tau(\rho)&=&\int d^2 x \rho \left(\pi -\tau \sqrt{|g|}\right)\\
  H(v)&=&\int d^2x \pi^{ab}(\mathcal L_v g)_{ab}.
 \end{array}
\end{equation}
The evolution with $H_{SD}$ of $\langle \pi \rangle$ is positive, so the ordering of this family of theories is in increasing York time $\tau$.

Let us conclude this section with the observation that for each  Fixed Shape Dynamics observable there is a representative function $O(\tau_1,\tau_2,p^1,p^2)$. In other words, the adapted phase space is given by the cotangent bundle over Teichm\"uller space.

\subsubsection{Formal Quantization}

The discussion of Dirac quantization of Shape Dynamics in \cite{Budd:2011er} is formal, because we considered formal wave functions $\psi[g]$, but abstained from specifying a kinematic inner product for these. We then proceeded by quantizing the local constraints of Shape Dynamics as vector fields on the metric configuration space and found that invariance under these vector fields implied that the wave function $\psi[g]=\psi(\tau_1,\tau_2;V)$ depends only on the Teichm\"uller parameters $\tau_1,\tau_2$ and the spatial volume $V$, which had to satisfy the Hamilton constraint
\begin{equation}
 \hat H_{SD} \psi_{SD}(\tau_1,\tau_2;V)=\left(-\tau_2^2(\partial^2_{\tau_1}+\partial^2_{\tau_2})+V^2(\partial_V^2+4\Lambda)\right)\psi(\tau_1,\tau_2;V)=0.
\end{equation}
Let us now apply the same strategy to Fixed Shape Dynamics: Following the strategy of the previous paragraph, we perform the same kinematic quantization, the only difference is that we will now replace the quantization of the Hamilton constraint with the quantization of $Q_o$, which is 
\begin{equation}
 \hat Q_o \psi_{FSD}(\tau_1,\tau_2;V)=(i \partial_V-\tau)\psi(\tau_1,\tau_2;V)=0. 
\end{equation}
We observe that the constraint $\hat Q_o$ is a first order differential equation in $V$, while the Shape Dynamics constraint $\hat H_{SD}$ is a second order differential equation in $V$, so there can never be a one to one identification of solutions. A possible solution to this problem goes as follows: solve the classically $H_{SD}=0$ for $\langle \pi \rangle=\pm\xi(\tau,p;V)=\pm\sqrt{4\Lambda+\frac{\tau_2^2}{V^2}((p^1)^2+(p^2)^2)}$ and observe that for $\Lambda>0$ there is an excluded range $-2\sqrt{\Lambda}<\langle \pi\rangle<2\sqrt{\Lambda}$ where no solutions to the constraint exists. This classical dichotomy can be used to motivate that only one sign is physical, since there are no trajectories connecting the two regimes. Following this reasoning one seeks a quantization $\hat \xi$ and replaces the original Shape Dynamics constraint with 
\begin{equation}
 i \partial_V \psi_{SD}(\tau_1,\tau_2;V)=\hat \xi \psi_{SD}(\tau_1,\tau_2;V).
\end{equation}
If we find a unitary time evolution operator $U(V,V_o)$ for this system, then we can choose an arbitrary but fix $V_o$ and compare the wave functions
\begin{equation}
 \begin{array}{rcl}
  \psi_{SD}(\tau_1,\tau_2;V)&=&U(V,V_o)\psi_o(\tau_1,\tau_2;V_o)\\
  \psi_{FSD}(\tau_1,\tau_2;V)&=&e^{-i\tau(V-V_o)} \psi_o(\tau_1,\tau_2;V_o).
 \end{array}
\end{equation}
It is a trivial observation that the matrix element of a Dirac observable of $U(V,V_o)OU(V_o,V)=O$ of Shape Dynamics is $\langle U(V,V_o)\psi^1_o(\tau_1,\tau_2;V_o),O U(V,V_o)\psi^2_o(\tau_1,\tau_2;V_o)\rangle$ is completely determined by the wave functions $\psi^1_o,\psi^2_o$ at $V_o$, so we can identify the observables of both theories with operators on $V$-independent wave functions $\psi(\tau_1,\tau_2;V_o)$. These operators are quantizations of the Fixed Shape Dynamics observables $O(\tau_1,\tau_2,p^1,p^2)$. This identification provides an explicit example where the reinterpretation of quantum Fixed Shape Dynamics observables as observables of quantum Shape Dynamics possible. This provides an explicit example where our our proposal of identifying Fixed Shape Dynamics matrix elements with matrix elements of Shape Dynamics as York time $\tau$ can be checked to work.

\subsection{General Quantization Strategy}

Let us now examine the general proposal of quantizing Fixed Shape Dynamics and reinterpreting the resulting theory as a quantization of General Relativity. This is of course only a formal discussion, because an explicit kinematic quantization of the gravity phase space exists presently only in the Loop Quantum Gravity framework which we reserve for a separate paper, due to technical subtleties. It is thus the purpose of this section to briefly discuss the the two-step process of first quantizing Fixed Shape Dynamics and then interpreting the resulting theory as a quantization of General Relativity through the classical observable equivalence of the two theories.

We start with a quantization of Fixed Shape Dynamics. The most direct way would be to perform a reduced phase space quantization on a Hilbert-space consisting of diffeomorphism-invariant Schr\"odinger wave functions $\psi[\rho]$, which may be obtained through Dirac quantization on a kinematic Hilbert-space $\mathcal K$ consisting of Schr\"odinger wave functions $\psi[g]$ that are square integrable w.r.t a diffeomorphism- and conformally- invariant measure $d\mu[g]$. One would then seek a kinematic quantization map $q$ that maps classical phase space functions $f[g,\pi]$ into essentially self-adjoint operators $q(f[g,\pi])$ on $\mathcal K$, that satisfies appropriate semiclassicality conditions. The next step consists of taking the quotient w.r.t. diffeomorphisms and conformal transformations. This results in a physical Hilbert space $\mathcal H$ and a quantization map $Q$ of conformal observables $O[\rho,\sigma]$ as essentially self-adjoint operators $Q(O(\rho,\sigma))$ on $\mathcal H$.

\paragraph{Problem of Time}

Supposing we have a found such a quantization of Fixed Shape Dynamics, then we we would seek to extract experimental predictions for General Relativity from the theory. We suggested above that we intend to interpret the expectation values $\langle v,Q(O)v\rangle$ as matrix elements for for the quantum gravity gravity observables $\langle v,Q(i(O))v\rangle$, where the denotes the identification of a Fixed Shape Dynamics observable $O[\rho,\sigma]$ with $O[P,\Sigma;\langle\pi\rangle]$. With this we have a natural physical interpretation of the observable algebra of quantum gravity coming from the duality of General Relativity and Shape Dynamics. The Fixed Shape Dynamics observable algebra coincides with the General Relativity observable algebra on $\Gamma_2(\rho,\sigma)$ and the matrix elements $\langle v,q(O)v\rangle$ have a natural interpretation as the expectation value of measurements of the Fixed Shape Dynamics observable $O$ in the quantum state $v$ and the spectra have the: 

This suggests the following experimental prescription for the interpretation of $\langle v,q(f(\rho,\sigma)) v\rangle$, where $v$ is the wave function of the universe: Consider the time slice with fixed York time $\tau$ and determine the value of the diffeomorphism-invariant functional $f(\rho,\sigma)$ in this time slice by your measurements. This interpretation is sticking with timlessness in the sense that physical measurements can only be performed at one particular value of York time. This means that then we have to find a physical interpretation of theory. This is certainly a hard foundational problem, that can not be attempted in this paper.

Let us now insist on the necessity for an evolution to contrast the timeless interpretation. For this one needs at least to be able to to relate two Fixed Shape Dynamics theories at different York times. The generator for this evolution between different Fixed Shape Dynamics theories is the standard Shape Dynamics Hamiltonian, which requires the general solution of the Lichnerowicz-York equation. One can practically use it only in those special situations where exact or approximate solutions to the Lichnerowicz-York equation are known (e.g. at large CMC-volume or when spatial derivatives are negligible). Moreover, implementing it as a unitary map between the Hilbert spaces of different Fixed Shape Dynamics seems thus only attainable approximately.

\section{Conclusions}\label{sec:Conclusions}

It is the purpose of this paper to provide a quantization strategy for the Wheeler-DeWitt equation by quantizing Shape Dynamics and using the classical equivalence between the ADM formulation of gravity and Shape Dynamics to reinterpret such a quantization as a quantization of gravity. This simple idea is complicated by two facts: (1) the Shape Dynamics Hamiltonian is nonlocal which complicates quantization considerably and (2) even if one is able to find the kernel of all Shape Dynamics constraints, one faces the problem of time. These two problems are not specific to this particular quantization strategy: problem (1) appeared in Isham's quantization program \cite{Isham:1992ms} for gravity in CMC gauge and the problem of time is one of the fundamental conceptual problems of the Wheeler-DeWitt program\cite{Anderson:2011jb,Barbour:2009zd}, which has been approached with various conceptual ideas, e.g. by Kucha\v{r}'s distinction between observables and perennials on the one hand and e.g. by timeless ``life inside an energy eigenstate'' \cite{Halliwell:2002th} on the other. 

A possible strategy, that is well adapted to the Wheeler-DeWitt program, is to quantize the timeless system but to provide enough structure to be able to recover time from the quantum system. This strategy does not solve the problem of time, but turns it into a problem of interpretation of the timeless quantum system. This said, we can state the two results of this paper:

(1) We introduced the notion of observable equivalence of gauge theories as a particular isomorphism of the observable algebras of two gauge theories. This notion of equivalence ``forgets'' about the time-ordering of classical trajectories, but solves the first problem, because it allows us to trade the nonlocal Shape Dynamics Hamiltonian for a one-parameter family of global conformal constraints. The resulting one parameter family of Fixed Shape Dynamics theories consists of completely local conformal theories. For each observable of Fixed Shape Dynamics there exists a preferred representative function that can easily be identified with a preferred representative of a Dirac-observable (or perennial) of ADM gravity by identifying these at fixed York time.

(2) We observed at the classical level that observable equivalence of a theory with time-reparametrization constraint with a one-parameter family of observable equivalent theories is sufficient to recover the original dynamics, because the time-ordering that is forgotten by observable equivalence can be recovered from an ordering of the one-parameter family of observable equivalent theories. Let us point out that this does not solve the problem of time, it only provides sufficient structure for its recovery at the classical level within the setting of equivalence of gauge theories.

To test the quantization proposal we considered pure gravity on a torus in 2+1 dimensions. The Shape Dynamics theory (or ADM gravity in CMC gauge) of this system can be quantized explicitly. We found that the standard quantization of Shape Dynamics leads to a constraint operator that is a second order differential equation in volume while the global conformal constraints of the one parameter family of Fixed Shape Dynamics theories are first order differential equation in volume. The two quantizations can thus not coincide. We then proposed a nonstandard quantization of Shape Dynamics (or ADM gravity in CMC gauge) for which the observable algebra can be argued to coincide with the observable algebra of Fixed Shape Dynamics at the quantum level. We thus have an example where the idea of quantizing Fixed Shape Dynamics and reinterpreting the resulting quantum theory as a quantization of Shape Dynamics (or ADM gravity in CMC gauge) can be identified with a particular quantization of Shape Dynamics.

\subsection*{Acknowledgements}

I wish to thank J. Barbour, S. Gryb and F. Mercati and in particular H. Gomes for stimulating discussions during our weekly hangouts. Research at the Perimeter Institute is supported in part by the Government of Canada through NSERC and by the Province of Ontario through MEDT.

\begin{appendix}

\section{Shape Dynamics Hamiltonian}

Previously, we constructed the Shape Dynamics Hamiltonian with an emphasis on the phase space reduction of the linking theory. It is however convenient for the present purpose to use a different construction, which makes use of standard arguments. We include these arguments for the sake of completeness.

The Shape Dynamics Hamiltonian can be determined through the dictionary theory, which is ADM-gravity in CMC gauge. This means that it is a representative of the CMC reparametrization constraint that is invariant under spatial volume preserving conformal transformations and spatial diffeomorphisms. To find this constraint one starts with the ADM-action, inserts the York decomposition ($\pi^{ab}=e^{-4\lambda}p^{ab}_{TT}+\frac \tau 2 \sqrt{|h|}h^{ab} e^{2\lambda}+\pi^{ab}_{long}$, $g_{ab}=e^{4\lambda}h_{ab}$, where $\pi^{ab}_{TT}$ is transverse-traceless w.r.t. $h_{ab}$ and $h_{ab}$ denotes a unique representative of each conformal equivalence class of metrics) of canonical variables with constant mean curvature and formally solves the constraint by setting the longitudinal part of the momenta to vanish $\pi^{ab}_{long}=0$ and the conformal factor to the solution of the Lichnerowicz-York equation $\lambda=\lambda_o$.
\begin{equation}
 \begin{array}{rcl}
  S_{ADM}&=&\int dt d^3x\left(\pi^{ab}\dot g_{ab}-S(N)-H(v)\right)\\
     &=&\int dt d^3x\left(e^{-4 \lambda_o} p^{ab}_{TT}+\frac 1 2 \tau \sqrt{|h|}h^{ab}\right)\left(4\dot \lambda_o e^{4\lambda_o}h_{ab}+e^{4\lambda_o}\dot h_{ab}\right)\\
     &=&\int dt d^3x\left(p^{ab}_{TT}\dot h_{ab}+\frac{d}{dt}(\tau e^{6\lambda_o}\sqrt{|h|})-\dot \tau e^{6\lambda_o}\sqrt{|h|}\right),
 \end{array}
\end{equation}
where we can disregard the boundary term $\frac{d}{dt}(\tau e^{6\lambda_o}\sqrt{|h|})$ and read-off the CMC reparametrization constraint as $p_\tau-\int d^3x e^{6\lambda_o} \sqrt{|h|}$. The representative function for $p_\tau$ that is invariant under spatial diffeomorphisms and volume-preserving conformal transformations and Poisson commutes with the adapted phase space variables $\rho_{ab},\sigma^{ab}$ is the total volume $V$, so we find the reparametrization constraint to be
\begin{equation}
 H_{SD}=V-\int d^3x e^{6\phi_o}\omega\approx 0,
\end{equation}
where $\phi_o[\rho,\sigma,\tau]$ denotes the solution to the Lichnerowicz-York equation. Invariance of $H_{SD}$ under spatial diffeomorphism is obvious and invariance under volume-preserving conformal transformations follows from that fact that of $\phi_o$ depends on $\sigma^{ab},\tau,h_{ab}$ only.

One can intuitively understand this as follows: The scalar constraints of ADM gravity are (on the Shape Dynamics constraint surface) equivalent to the condition that the conformal factor satisfies the Lichnerowicz-York equation. The total volume is the only conformal degree of freedom that is not pure gauge, so one can understand that the leftover scalar constraint can be locally solved for the volume and uniqueness \cite{YorkOMur} of the positive solution to the Lichnerowicz-York equation provides that the volume constraint is equivalent to the leftover scalar constraint.

\end{appendix}

\end{document}